\documentclass[a4paper]{llncs}
\usepackage{graphicx}
\usepackage{amsfonts}
\usepackage{multirow}
\usepackage{mathtools}
\usepackage{enumitem}
\usepackage{sidecap}
\usepackage{bm}

\usepackage{shortcuts_jac_miccai}

\usepackage{numprint} % for spaces at 10^3, 10^6 etc. for large numbers
\npthousandsep{\,}

\def\*#1{\mathbf{#1}}

\begin{document}
\title{Statistical Inference with Ensemble of Clustered Desparsified Lasso}
\titlerunning{Statistical Inference with Ensemble of Clustered Desparsified Lasso}
% \author{Anonymous Authors}
% \institute{Anonymous Institute}

\author{J\'er\^ome-Alexis \textsc{Chevalier}\inst{1, 2}
\and Joseph \textsc{Salmon}\inst{2}
\and Bertrand \textsc{Thirion}\inst{1}}
% index{Chevalier, J\'er\^ome-Alexis}
% index{Salmon, Joseph}
% index{Thirion, Bertrand}
\institute{
Parietal Team, Inria, CEA, Paris-Saclay University, France
\and
Telecom ParisTech, Paris, France\\
\email{jerome-alexis.chevalier@inria.fr}}

\maketitle
\sloppy
\begin{abstract}
Medical imaging involves high-dimensional data, yet their acquisition is obtained for limited samples.
Multivariate predictive models have become popular in the last decades to fit some external variables from imaging data, and standard algorithms yield point estimates of the model parameters.
It is however challenging to attribute confidence to these parameter estimates, which makes solutions hardly trustworthy.
In this paper we present a new algorithm that assesses parameters statistical significance and that can scale even when the number of predictors $p \geq 10^{5}$ is much higher than the number of samples $n \leq 10^{3}$, by leveraging structure among features.
Our algorithm combines three main ingredients: a powerful inference procedure for linear models --the so-called Desparsified Lasso-- feature clustering and an ensembling step.
We first establish that Desparsified Lasso alone cannot handle $n \ll p$ regimes; then we demonstrate that the combination of clustering and ensembling provides an accurate solution, whose specificity is controlled.
We also demonstrate stability improvements on two neuroimaging datasets.
\end{abstract}
%
%%%%%%%%%%%%%%%%%%%%%%%%%%%%%%%%%%%%%%%%%%%%%%%%%%%%%%%%%%%%%%%%%%%%%%%%%%%%%%%
%%%%%%%%%%%%%%%%%%%%%%%%%%%%%%%%%%%%%%%%%%%%%%%%%%%%%%%%%%%%%%%%%%%%%%%%%%%%%%%
\section{Introduction}
\label{sec:introduction}
%%%%%%%%%%%%%%%%%%%%%%%%%%%%%%%%%%%%%%%%%%%%%%%%%%%%%%%%%%%%%%%%%%%%%%%%%%%%%%%
%%%%%%%%%%%%%%%%%%%%%%%%%%%%%%%%%%%%%%%%%%%%%%%%%%%%%%%%%%%%%%%%%%%%%%%%%%%%%%%
%
Prediction problems in medical imaging are typically high-dimensional small-sample problems.
Training such models can be seen as an inference procedure.
As in all research fields, this inference has to come with probabilistic guarantees in order to assess its reliability and to clarify further interpretation.
In such settings, linear models have raised a strong interest. In particular  the Lasso, introduced in \cite{Tibshirani94}, has been thoroughly investigated in \cite{Hastie_Tibshirani_Wainwright15}.
Specifically, for settings in which the number of features $p$ is greater than the number of samples $n$ --though commensurate-- numerous inference solutions have been proposed: see among others
\cite{buhlmann2013,Dezeure2015,Meinshausen2008,wasserman2009,Zhang_Zhang14}.
However, when $n \ll p$, these inference solutions are not scalable.
In practice they fail to be informative, as we will show in our first simulation (\lcf \Cref{sub:first_simu_clustering}).
Indeed, in these regimes, due to the curse of dimensionality, localizing statistical effects becomes much harder and an informative inference seems hopeless without dimensionality reduction.
However, in high dimension, datasets often exhibit a particular data structure and inter-predictors correlation.
This makes dimension reduction possible by the means of clustering algorithms which should respect data structure as described in \cite{Varoquaux2012}.
The issue with clustering-based solutions is that clustering is almost surely suboptimal and unstable; it carries some arbitrariness related to initialization or estimators heuristics.
A solution to mitigate this confounding factor is to embed it in a bagging strategy, as done \eg in \cite{Varoquaux2012}.\\\indent
Our contribution is an algorithm for statistical inference in high-dimensional scenarios, combining the \DL procedure, first introduced in \cite{Zhang_Zhang14}, with clustering and bagging steps: the Ensemble of Clustered \DL.
We describe it in detail and provide experiments on simulated and real data to assess its potential on multivariate linear models for medical imaging.
%
%%%%%%%%%%%%%%%%%%%%%%%%%%%%%%%%%%%%%%%%%%%%%%%%%%%%%%%%%%%%%%%%%%%%%%%%%%%%%%%
%%%%%%%%%%%%%%%%%%%%%%%%%%%%%%%%%%%%%%%%%%%%%%%%%%%%%%%%%%%%%%%%%%%%%%%%%%%%%%%
\section{Theoretical and Algorithmic Framework}
\label{sec:theoretical_and_algorithmic_framework}
%%%%%%%%%%%%%%%%%%%%%%%%%%%%%%%%%%%%%%%%%%%%%%%%%%%%%%%%%%%%%%%%%%%%%%%%%%%%%%%
%%%%%%%%%%%%%%%%%%%%%%%%%%%%%%%%%%%%%%%%%%%%%%%%%%%%%%%%%%%%%%%%%%%%%%%%%%%%%%%
%
\paragraph{\textbf{Notations.}}
For clarity, scalars are denoted with normal font, vectors with bold lowercase, and matrices with bold uppercase letters.
For $p \in \bbN, [p] = \discset{1, \ldots, p}$.
\paragraph{\textbf{Inference on Linear Models.}}
Our aim is to give confidence bounds on the coefficients of the parameter vector denoted $\*w^*$ in the following linear model:
\begin{equation}\label{eq:noise_model}
\*y = \*X\*w^* + \sigma_{*} \bm\varepsilon \enspace,
\end{equation}
where $\*y \in \bbR^{n}, \*X \in \bbR^{n \times p}, \*w^* \in \bbR^{p}$, $\bm\varepsilon \sim \mathcal{N}(\*0,\*I_n)$ and $\sigma_*>0$.
The matrix $\*X$ is the design matrix, its columns are called the predictors, $\*y$ is called the response vector, $\bm\varepsilon$ is the noise (or the error vector) of the model and $\sigma_{*}$ is the (unknown) noise standard deviation.
The signal to noise ratio (SNR),  defined by $\SNR_{y} = \normin{\*X\*w^*}_2 / (\sigma_*\normin{\bm\epsilon}_2)$, is a measure that describes the noise regime in any given experiment.
The true support is defined as $S_* = \discsetin{j \in [p];w^*_j \neq 0}$ and its size is $s_* = |S_*|$.
It is noteworthy that our problem is not a prediction problem, we are not aiming at finding $\hat{\*w}$ minimizing $\normin{\*X \hat{\*w} - \*X \*w^*}_2$, but an estimation problem, in which we want to control $\|\hat{\*w} - \*w^*\|_\infty$ statistically.
\paragraph{\textbf{\DL for High-Dimensional Inference.}}
The \DL (DL) estimator denoted $\hat{\*w}^{\rm{DL}}$, introduced in \cite{Zhang_Zhang14}, can be seen as a generalization of the Ordinary Least Squares (OLS) estimator for inference in $n < p$ settings.
Under some assumptions (notably the sparsity of $\*w^*$) that are made explicit in \cite{Dezeure2015}, $\hat{\*w}^{\rm{DL}}$ has the following property:
\begin{equation}\label{eq:DL_cb}
\forall j \in  [p], \quad \sigma^{-1}_{*}(\Omega_{jj})^{-1/2}(\hat{w}^{\rm{DL}}_j-w^*_j) \sim \mathcal{N}(0,1) \enspace,
\end{equation}
where the diagonal of $\bm\Omega$ (estimated precision matrix; inverse of $\*X^\top \*X / n$) is computed concurrently with $\hat{\*w}^{\rm{DL}}$, as described in \cite{Zhang_Zhang14}.
From \Cref{eq:DL_cb} we can compute confidence intervals and p-values of the coefficients of the weight vector.\\\indent
In \cite{Dezeure2015}, several high-dimensional inference solutions are discussed and compared.
DL displays an interesting trade-off between good control of the family-wise error rate (FWER) and strong power.
The FWER is defined as $\mbox{FWER} = \mbox{Prob}(\mbox{FP} \ge 1)$ where FP is the number of false positives.
\paragraph{\textbf{Clustering to Handle Structured High-Dimensional Data.}}
In high-dimensional inference, variables are often highly correlated.
Specifically, a medical image has a 3D representation and a given voxel is highly correlated with its neighbors; $\*w$ obviously carries the same structure.
In addition $n \ll p$ and $n < s_*$ make the statistical inference challenging without data structure assumptions as shown in \cite{Dezeure2015}.
To leverage data structure, we introduce a clustering step that reduces data dimensionality before applying our inference procedure.
Here, we consider a spatially-constrained hierarchical clustering algorithm described in \cite{Varoquaux2012} that uses Ward criterion while respecting the image geometrical structure.
The combination of this clustering algorithm and the DL inference procedure will be referred to as the Clustered \DL (CDL) algorithm.
\paragraph{\textbf{Bagging to Alleviate Dependency on Clustering.}}
It is preferable not to rely on a particular clustering as small perturbations on it have a dramatic impact on the final solution.
We followed the approach presented in \cite{Varoquaux2012} that argues in favor of the randomization over a spatially-constrained clustering method: to build $B$ clusterings of the predictors, they use the same clustering method but with $B$ different random subsamples of size $\lfloor{0.7 n}\rfloor$ from the full sample.
\paragraph{\textbf{Inference with Ensemble of Clustered \DL.}}
\begin{figure}[!t]
    \includegraphics[width=0.95\linewidth]{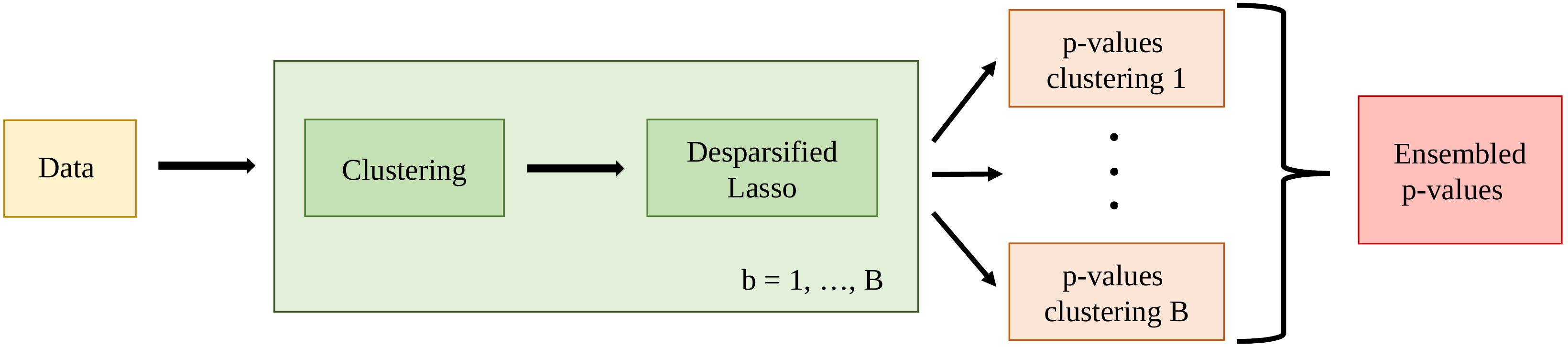}
    \caption{\label{fig:algorithm_diagram}
    To leverage \DL inference procedure on medical images, our algorithm relies on feature clustering and an ensembling step on randomized solutions.
}
\end{figure}
We now have all the elements to present our Ensemble of Clustered \DL (ECDL) algorithm which is summarized in \Cref{fig:algorithm_diagram}.
ECDL consists in $B$ repetitions of the CDL algorithm (using random subsamples of size $\lfloor{0.7 n}\rfloor$ for the clustering and the full sample for the DL procedure) and an ensembling step analogous to the bagging method introduced by \cite{Breiman1996}.
Once the CDL algorithm has been run $B$ times, we have $B$ partitions into $C$ clusters, each cluster being associated with a p-value.
We denote by $P^{(b, c)}$ the p-value for the $c^{th}$ cluster in the $b^{th}$ fold.
The p-value $P^{(b)}_j$ of the coefficient $j \in [p]$ in the $b^{th}$ repetition is $P^{(b, c)}$ whenever $j$ belongs to cluster $c$, \ie we attribute the same p-value to all the predictors in a given cluster.
This yields $B$ p-values for each coefficient.
Finally, to ensemble the p-values one has to use specific techniques which ensure that the resulting p-value is meaningful as a frequentist hypothesis test.
Thus, to derive the p-value $P_j$ of the $j^{th}$ coefficient, we have considered the ensembling solution presented in \cite{Meinshausen2008} that has the required properties and consists in taking the median of $\{ P^{(b)}_j ~ \mbox{for} ~ b \in [B] \} $ multiplied by $2$.
%
%%%%%%%%%%%%%%%%%%%%%%%%%%%%%%%%%%%%%%%%%%%%%%%%%%%%%%%%%%%%%%%%%%%%%%%%%%%%%%%
%%%%%%%%%%%%%%%%%%%%%%%%%%%%%%%%%%%%%%%%%%%%%%%%%%%%%%%%%%%%%%%%%%%%%%%%%%%%%%%
\section{Simulation and Experimental Results}
\label{sec:simulation_and_experimental_results}
%%%%%%%%%%%%%%%%%%%%%%%%%%%%%%%%%%%%%%%%%%%%%%%%%%%%%%%%%%%%%%%%%%%%%%%%%%%%%%%
%%%%%%%%%%%%%%%%%%%%%%%%%%%%%%%%%%%%%%%%%%%%%%%%%%%%%%%%%%%%%%%%%%%%%%%%%%%%%%%
%
%%%%%%%%%%%%%%%%%%%%%%%%%%%%%%%%%%%%%%%%%%%%%%%%%%%%%%%%%%%%%%%%%%%%%%%%%%%%%%%
\subsection{First Simulation: the Importance of Dimension Reduction}
\label{sub:first_simu_clustering}
%%%%%%%%%%%%%%%%%%%%%%%%%%%%%%%%%%%%%%%%%%%%%%%%%%%%%%%%%%%%%%%%%%%%%%%%%%%%%%%
%
\paragraph{\textbf{Simulation.}}
This simulation has a 1D structure and we set $n = 100$ and $p = \numprint{2000}$.
We construct the design matrix $\*X$ such that predictors are normally distributed and two consecutive predictors have a fixed correlation $\rho = 0.95$.
The weight $\*w^*$ is such that $w^*_j = 1$ for $1 \leq j \leq 50$ and $w^*_j = 0$ otherwise, then $s_{*} = 50$.
We also set $\sigma_{*} = 10$ such that $\mbox{SNR}_{y} = 3$ (\lcf \Cref{sub:experiments_mri}).
\begin{figure}[!b]
    \centering
    \begin{minipage}[c]{.41\linewidth}
        \centering
	    \includegraphics[width=\linewidth]
    	{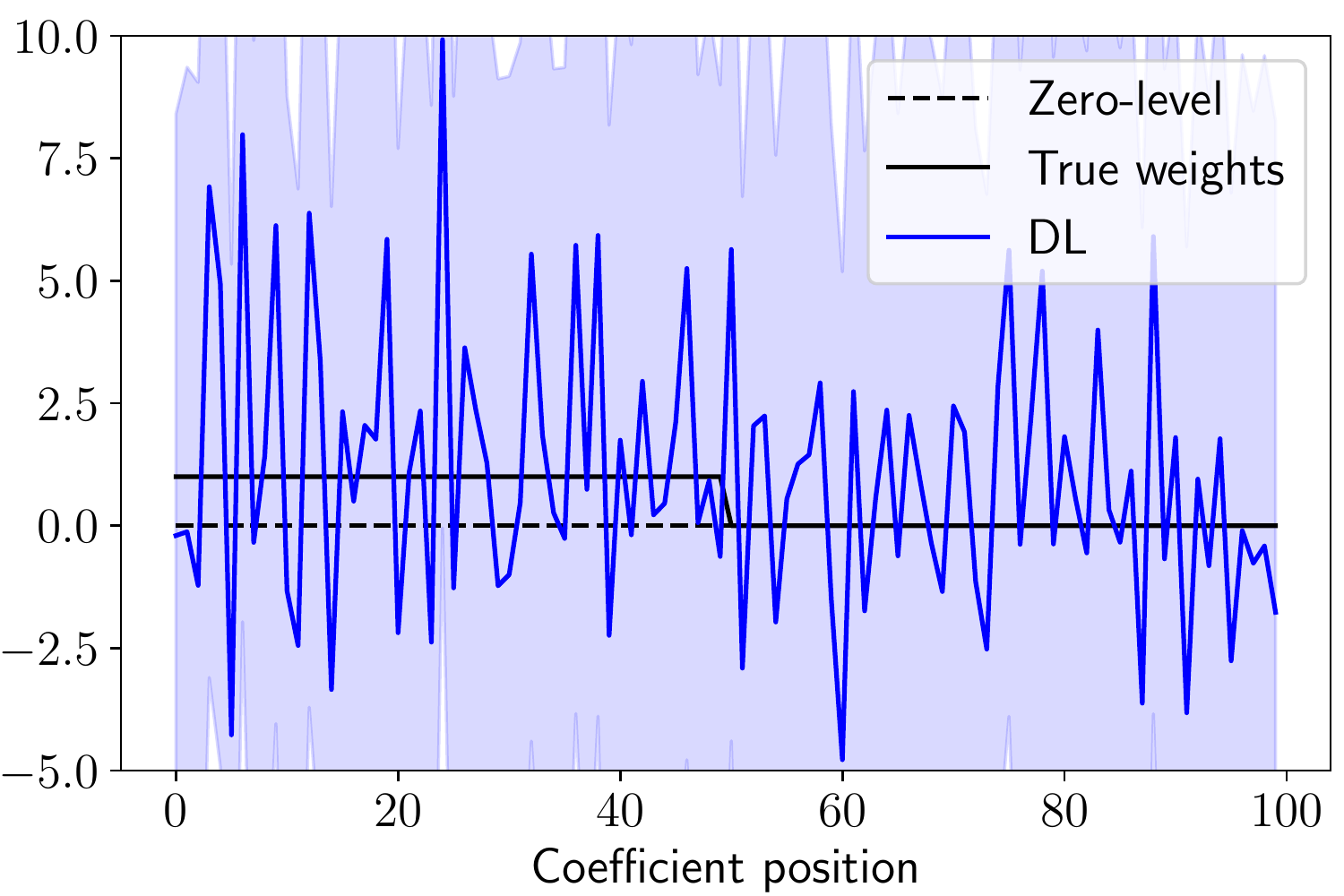}
    	\\(a)
    \end{minipage}
    \begin{minipage}[c]{.41\linewidth}
        \centering
        \includegraphics[width=\linewidth]
        {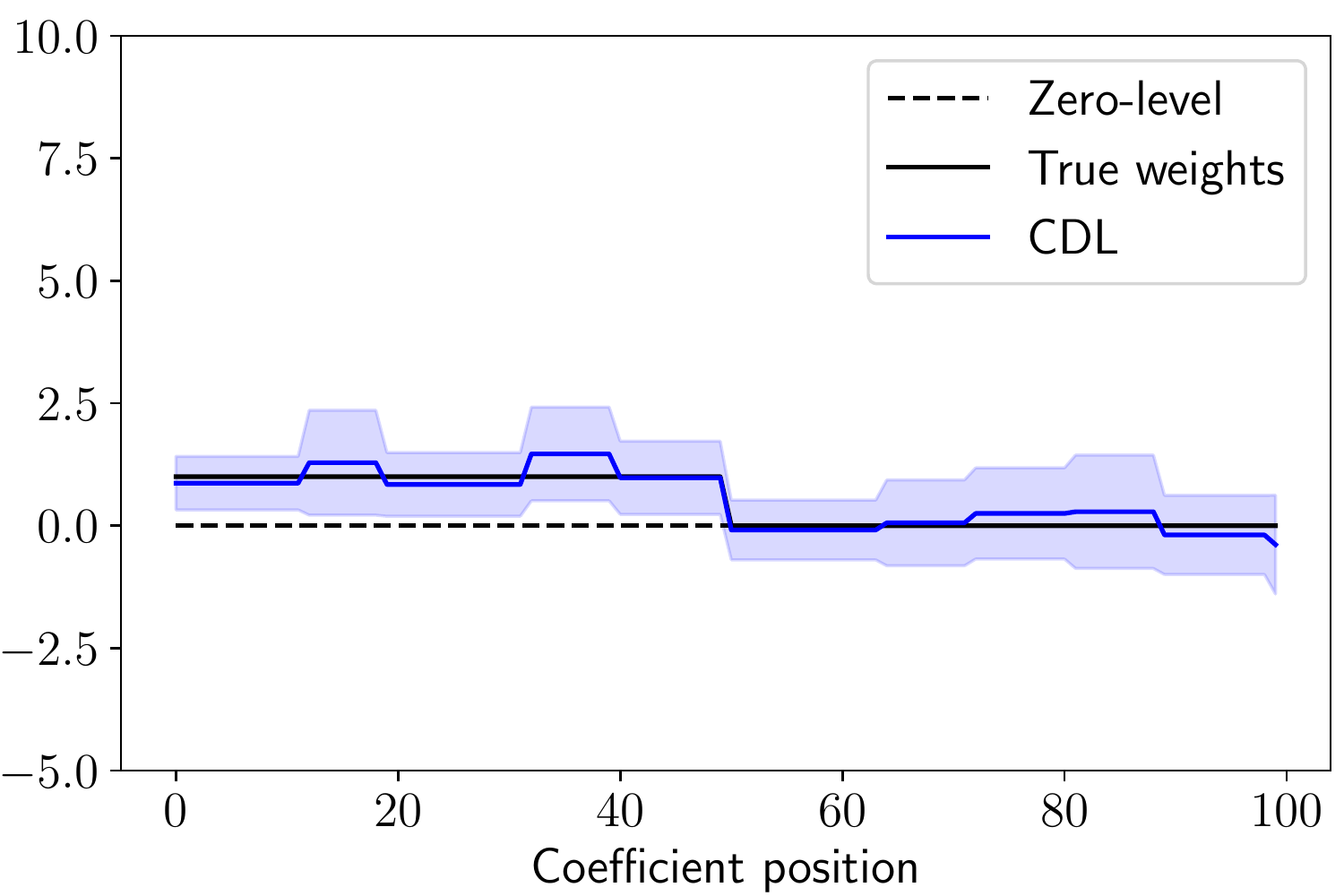}
        \\(b)
    \end{minipage}
    \caption{\label{fig:simu1}
    (a) 95\% coefficient intervals given by the raw \DL (DL) fail to retrieve the true support.
	  (b) 95\% coefficient intervals given by the Clustered \DL
    (CDL) are much narrower, and yield a good support accurately.
	}
\end{figure}
\paragraph{\textbf{Results.}}
We compare the DL procedure applied to the uncompressed data, displayed in \Cref{fig:simu1}-(a), and the CDL algorithm in \Cref{fig:simu1}-(b).
The number of clusters, whose impact will be discussed in \Cref{sec:discussion}, has been set to $C = 200$, allowing to reduce the dimension from $p = \numprint{2000}$ to $C = 200$ before performing the inference.
This reduction tames the estimator variance and yields useful confidence intervals that could not be reached by DL only.
%
%%%%%%%%%%%%%%%%%%%%%%%%%%%%%%%%%%%%%%%%%%%%%%%%%%%%%%%%%%%%%%%%%%%%%%%%%%%%%%%
\subsection{Second Simulation: Improvement by Bootstrap and Aggregation}
\label{sub:second_simu_boostrap_aggregation}
\begin{figure}[!t]
    \centering
    \begin{minipage}[c]{.305\linewidth}
      \centering
	    \includegraphics[width=\linewidth]
    	{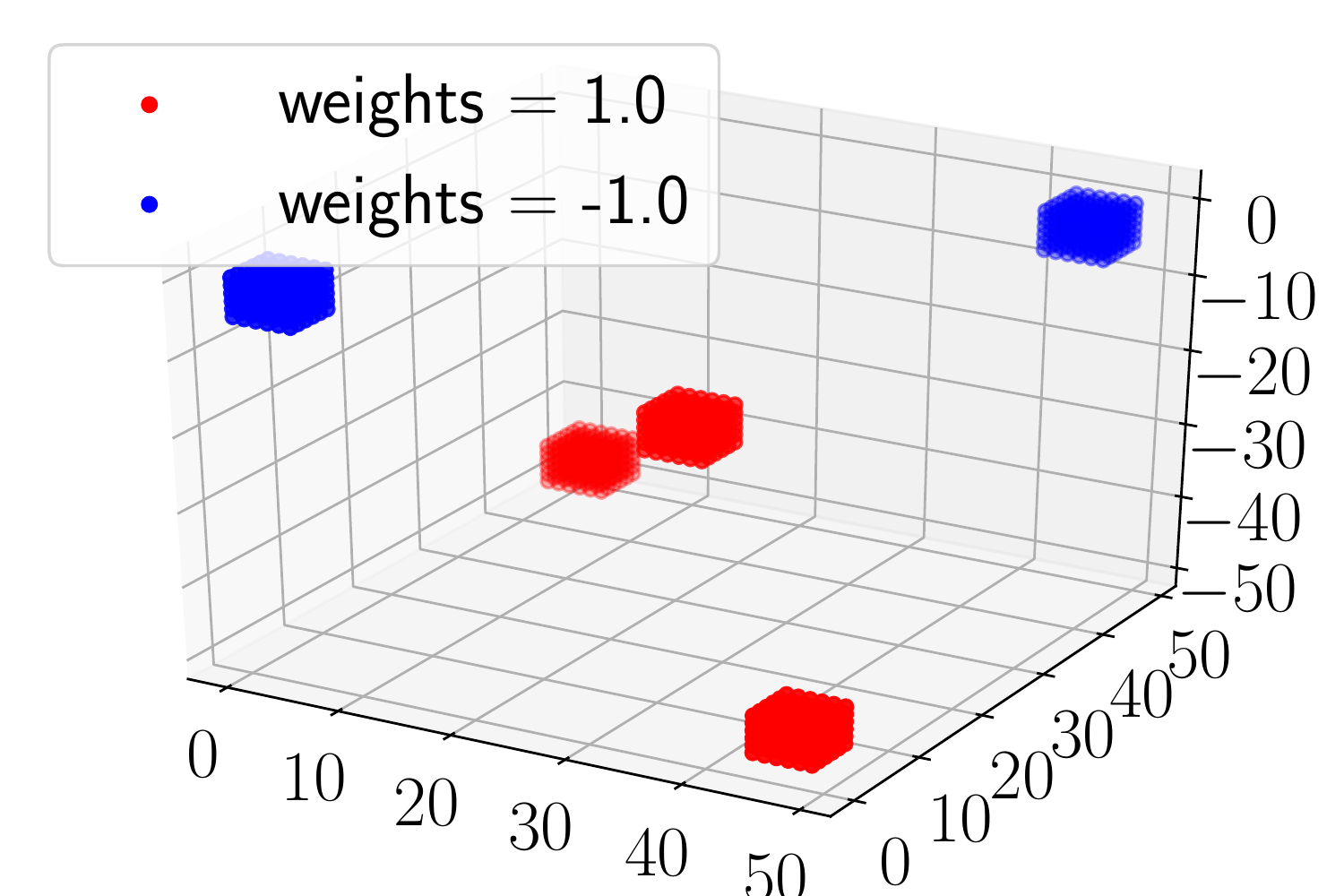}
    	\\(a) 3D weight vector: $\tilde{\*w}^*$
    \end{minipage}
    \begin{minipage}[c]{.305\linewidth}
      \centering
	    \includegraphics[width=\linewidth]
    	{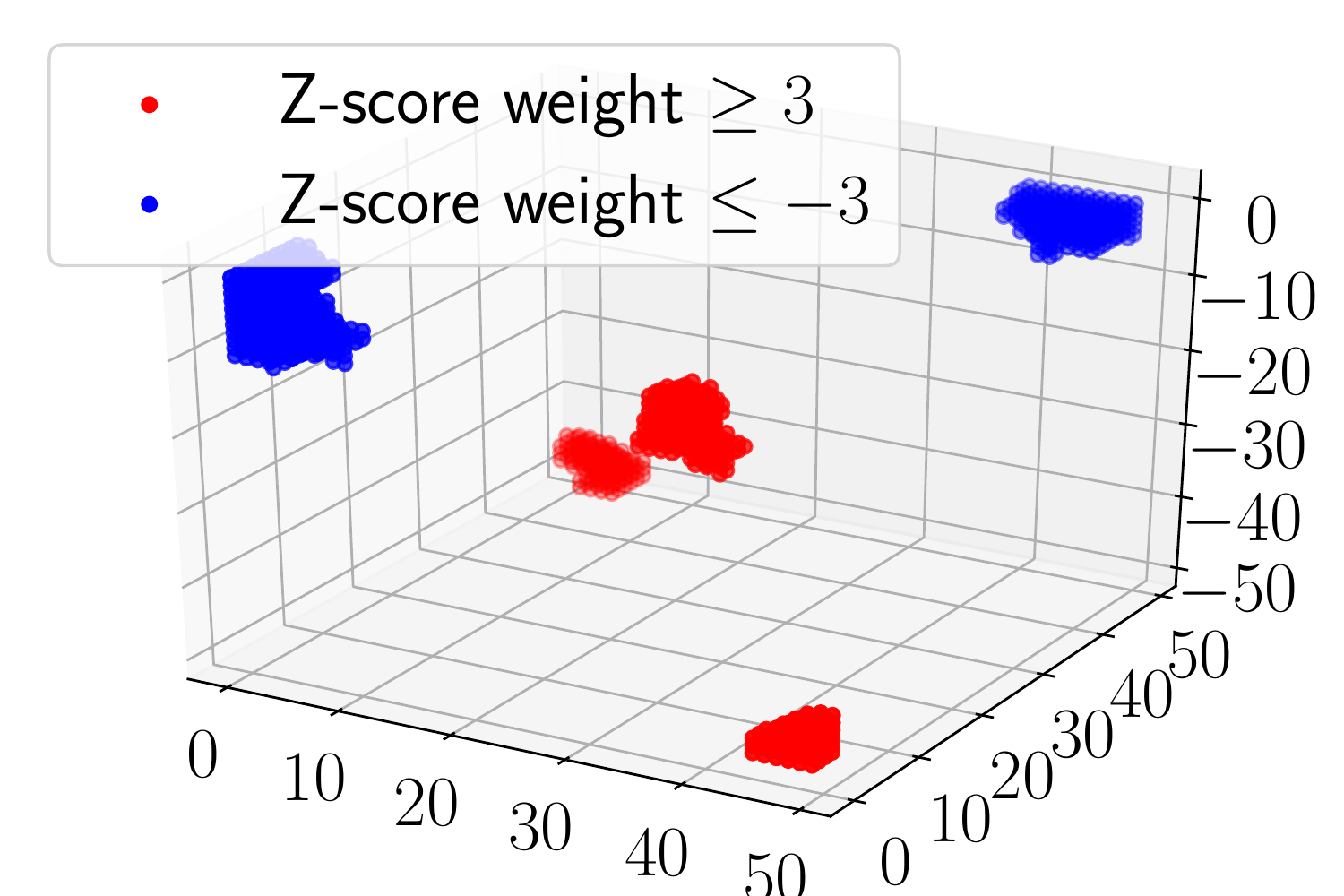}
    	\\(b) CDL
    \end{minipage}
    \begin{minipage}[c]{.305\linewidth}
      \centering
	    \includegraphics[width=\linewidth]
    	{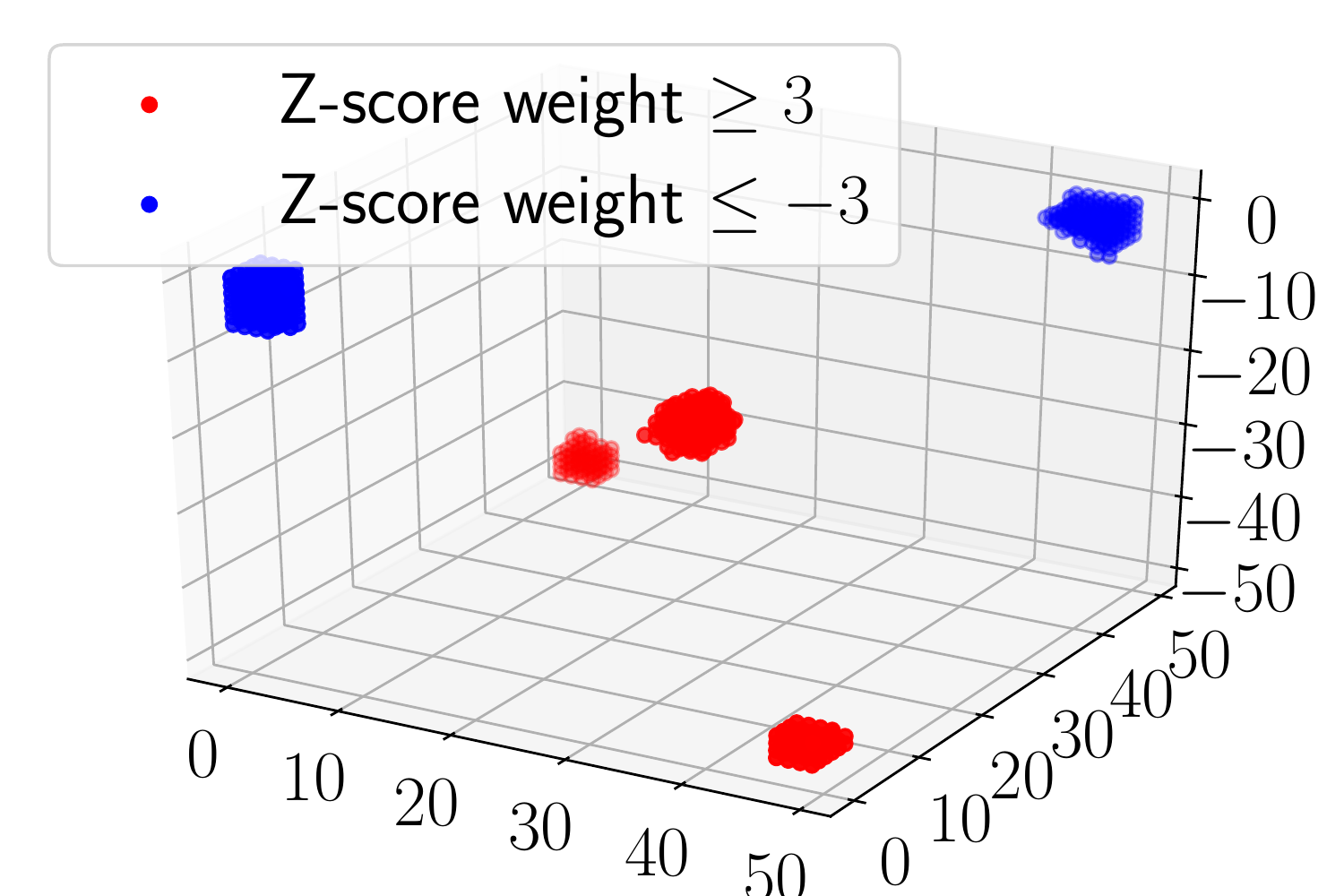}
    	\\(c) ECDL
    \end{minipage}
    \caption{\label{fig:weight_vector}
    In this simulation, comparing the original 3D weight vector with CDL and ECDL solutions, we observe that the ECDL solution is much more accurate.
	}
\end{figure}

%%%%%%%%%%%%%%%%%%%%%%%%%%%%%%%%%%%%%%%%%%%%%%%%%%%%%%%%%%%%%%%%%%%%%%%%%%%%%%%
%
\paragraph{\textbf{Simulation.}}
Here, we consider a simulation with a 3D structure, that aims at approximating the statistics of the Oasis experiment (\lcf \Cref{sub:experiments_mri}).
The volume considered is a 3D-cube with edge length $H = 50$, with $n = 400$ samples and $p = H^{3} = \numprint{125000}$ predictors (voxels).
To construct $\*w^*$, we define a 3D weight vector $\tilde{\*w}^*$ with five regions of interest (ROIs) represented in \Cref{fig:weight_vector}-(a) and then make a bijective transformation of $\tilde{\*w}^*$ in a vector of size $p$.
Each ROI is a cube of width $h = 6$, leading to a size of support $s_{*} = 5 h^{3} = \numprint{1080}$.
Four ROIs are situated in corners of the cubic map and the last ROI is situated in the center of the cube.
To construct $\*X$, we first define the 3D design matrix $\tilde{\*X}$ from $p$ random normal vectors of size $n$ smoothed with 3D Gaussian filter with bandwidth $\sigma_{\rm{smth}}$ (smoothing is performed across all predictors for each sample), then we use the same transformation as before and derive the $n \times p$ design matrix.
The choice $\sigma_{\rm{smth}}=2$ is made to achieve similar correlations as for the Oasis experiment.
We also set $\sigma_{*} = 8$, \ie $\mbox{SNR}_{y} = 3$ (\lcf \Cref{sub:experiments_mri}).
\begin{figure}[!b]
    \centering
    \begin{minipage}[c]{.41\linewidth}
        \centering
	    \includegraphics[width=\linewidth]
    	{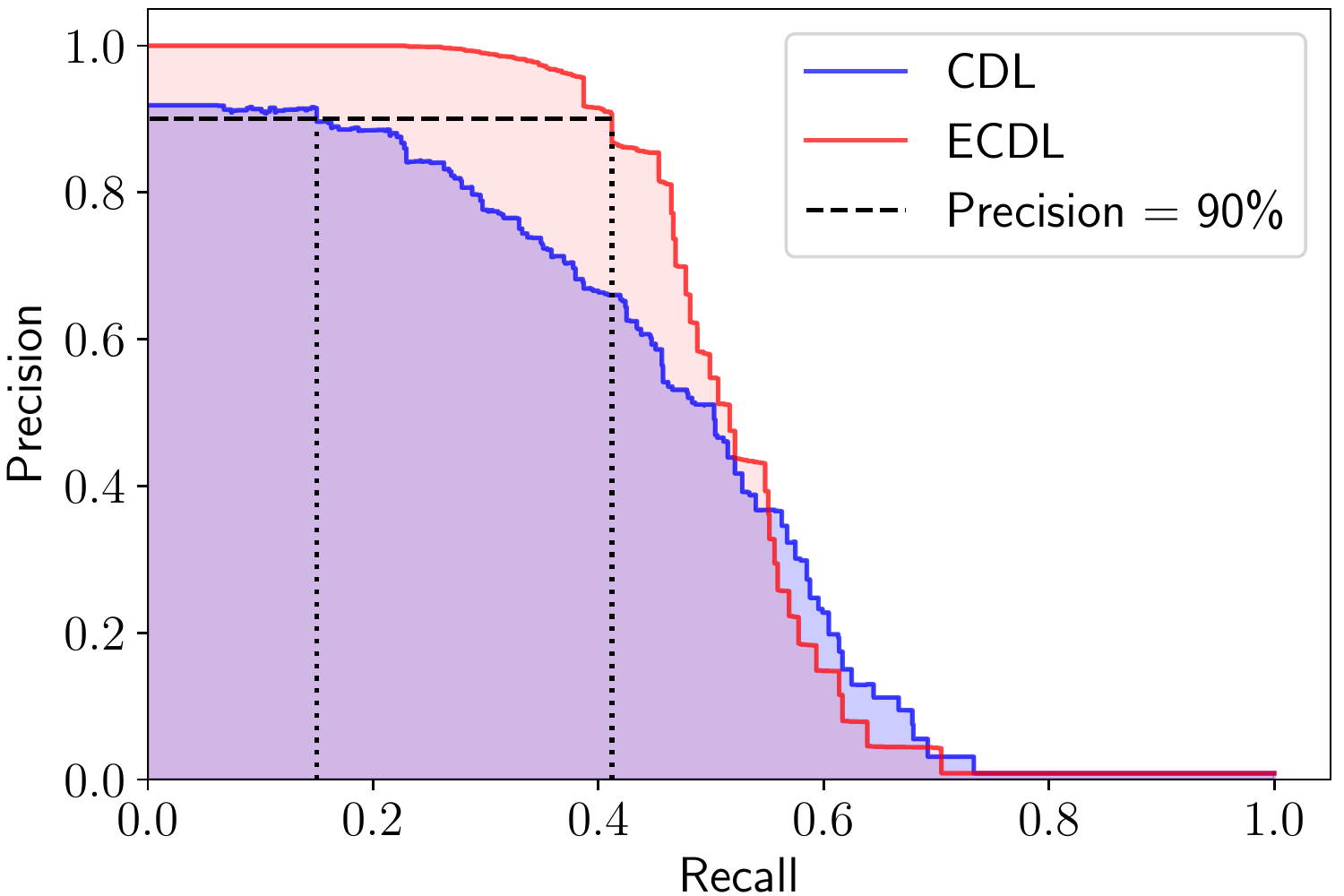}
    	\\(a)
    \end{minipage}
    \begin{minipage}[c]{.41\linewidth}
        \centering
        \includegraphics[width=\linewidth]
        {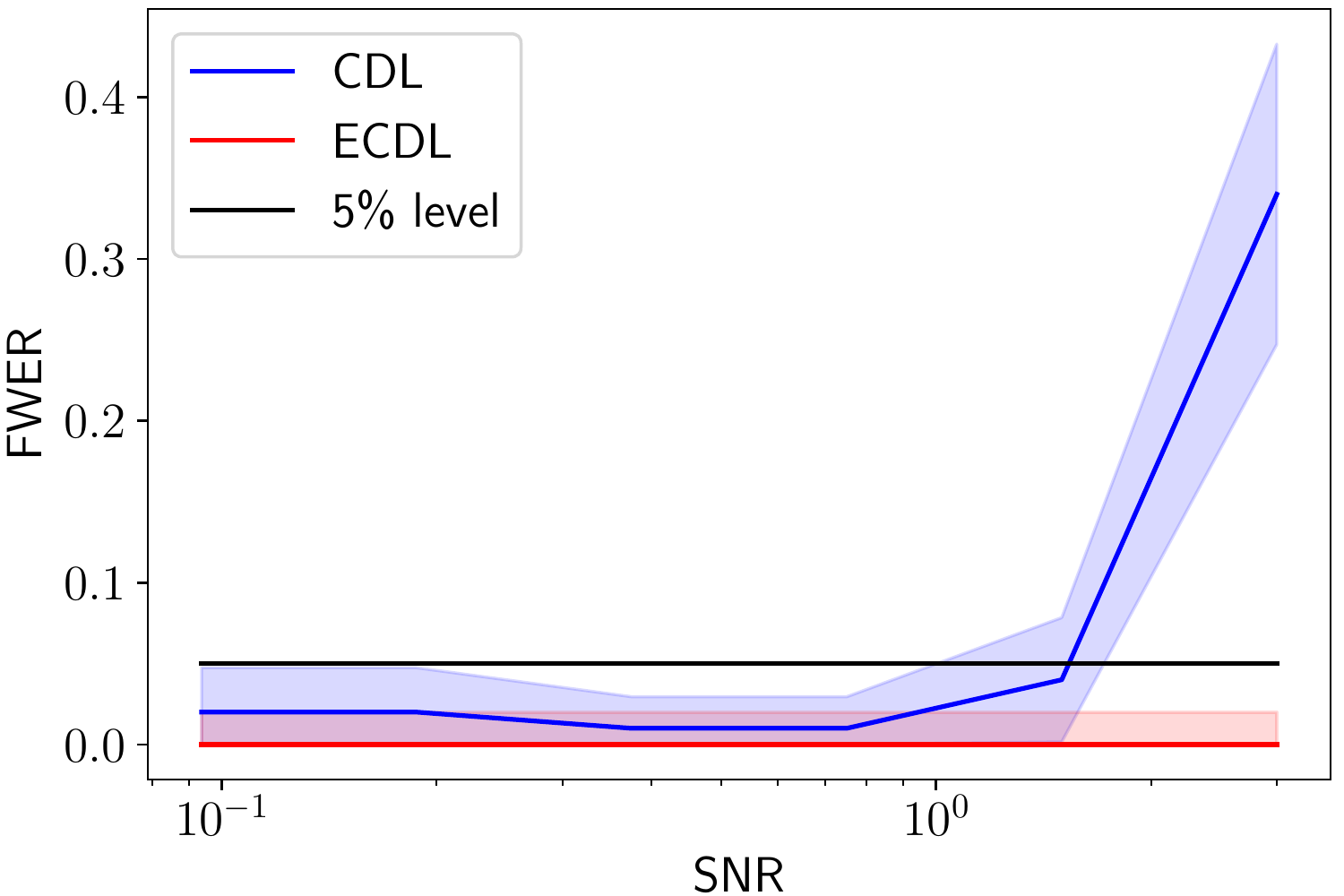}
        \\(b)
    \end{minipage}
    \caption{\label{fig:simu2}
    (a) The precision-recall curve for the recovery of $\*w$ is much better adding an ensembling step over CDL.
	  (b) FWER (nominal rate 5\%) is well controlled by the ECDL algorithm while for high level of SNR it is not controlled by the CDL algorithm.
	}
\end{figure}
\paragraph{\textbf{Results.}}
To derive the ECDL solutions we aggregated $B = 25$ different CDL solutions during the ensembling step.
To obtain the results presented in \Cref{fig:simu2}, we ran $100$ simulations.
In \Cref{fig:simu2}-(a), we display the precision-recall curve (\lcf Scikit-learn {\tt precision\_recall\_curve} function) of the solutions obtained by each algorithm with $C = 500$ clusters.
ECDL very strongly outperforms CDL: for precision of at least $90\%$, the ECDL recall is $42\%$ while the CDL recall is only $16\%$.\\\indent
In order to check the FWER control, we define a neutral region that separates ROIs from the non-active region.
Indeed, since the predictors are highly correlated, the detection of a null predictor in the vicinity of an active one is not a mistake.
Thus, neutral regions enfold ROIs with a margin of 5 voxels.
We compare different values of $\sigma_{*}$ from $2^{3}$ to $2^{8}$ giving $\mbox{SNR}_{y}$ lying between $.1$ and $3$.
In \Cref{fig:simu2}-(b), one can observe that the FWER is always well controlled using ECDL; the later is even conservative since the empirical FWER stays at $0\%$ for a $5\%$ nominal level.
On the opposite, the FWER is not well controlled by CDL: its empirical value goes far above the $5\%$ rate for high SNR.
This is due to the shape of the discovered regions that do not always correspond to the exact shape and location of ROIs.
This effect is also observable watching thresholded Z-score maps yielded by CDL and ECDL in \Cref{fig:weight_vector}.
By increasing the number of clusters, we would obtain discovered regions more similar to the true ROIs, yet their statistical significance would drop and the power would collapse.
%
%%%%%%%%%%%%%%%%%%%%%%%%%%%%%%%%%%%%%%%%%%%%%%%%%%%%%%%%%%%%%%%%%%%%%%%%%%%%%%%
\subsection{Experiments on MRI Datasets}
\label{sub:experiments_mri}
%%%%%%%%%%%%%%%%%%%%%%%%%%%%%%%%%%%%%%%%%%%%%%%%%%%%%%%%%%%%%%%%%%%%%%%%%%%%%%%
%
\paragraph{\textbf{Haxby Dataset.}}
Haxby is a functional MRI dataset that maps the brain responses of subjects watching images of different objects (see \cite{Haxby2001}).
In our study we only consider the responses related to images of faces and houses for the first subject, to identify brain regions that  discriminate between these two stimuli, assuming that this problem can be modeled as a regression problem.
Here $n = 200$, $p = \numprint{24}$k, (estimated) $\mbox{SNR}_{y} = 1.0$ and we used $C = 500$ and $B = 25$.
\paragraph{\textbf{Oasis Dataset.}}
The Oasis MRI dataset (see \cite{Marcus2007}) provides anatomical brain images of several subjects together with their age.
The SPM voxel-based morphometry pipeline was used to obtain individual gray matter density maps.
We aim at identifying which regions are informative to predict the age of a given subject.
Here $n = 400$, $p = \numprint{125}$k and (estimated) $\mbox{SNR}_{y} = 3.0$; we also took $C = 500$ and $B = 25$ as in \Cref{sub:second_simu_boostrap_aggregation}.
\begin{figure}[!b]
    \centering
    \begin{minipage}[c]{.18\linewidth}
        \centering
	    \includegraphics[width=\linewidth]
    	{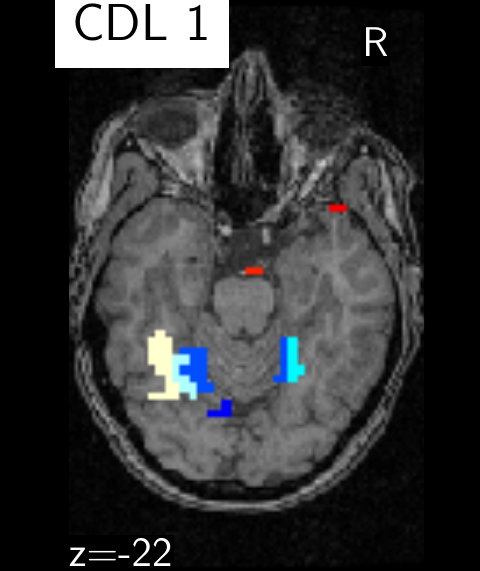}
    \end{minipage}
    \begin{minipage}[c]{.18\linewidth}
        \centering
        \includegraphics[width=\linewidth]
        {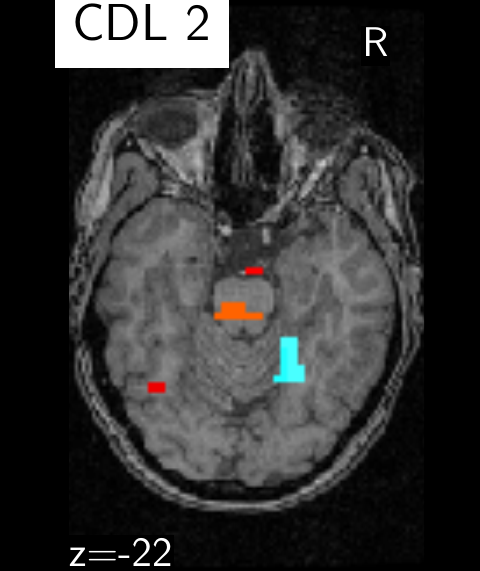}
    \end{minipage}
    \begin{minipage}[c]{.18\linewidth}
        \centering
        \includegraphics[width=\linewidth]
        {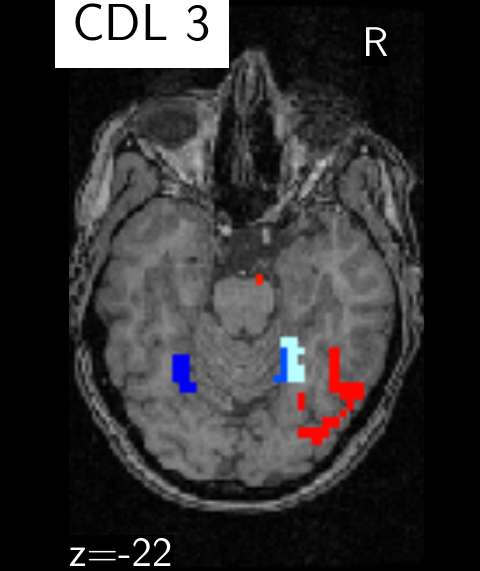}
    \end{minipage}
    \hspace{5mm}
    \begin{minipage}[c]{.27\linewidth}
        \centering
        \includegraphics[width=\linewidth]
        {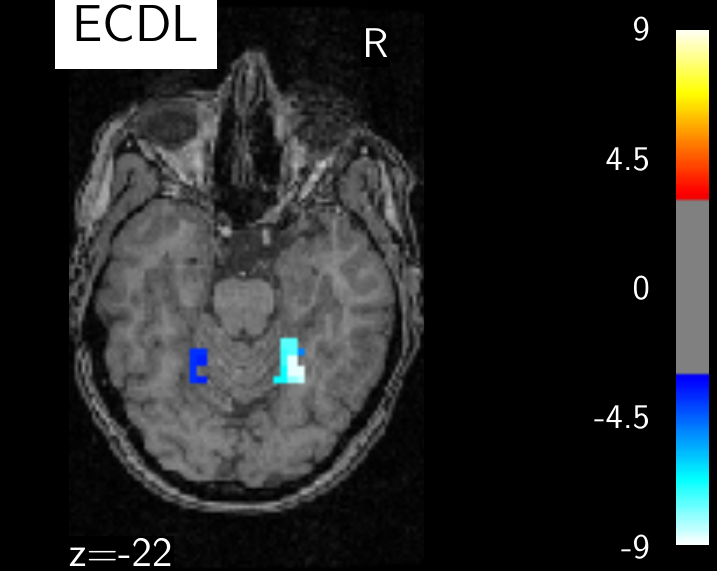}
    \end{minipage}

    \begin{minipage}[c]{.18\linewidth}
        \centering
      \includegraphics[width=\linewidth]
      {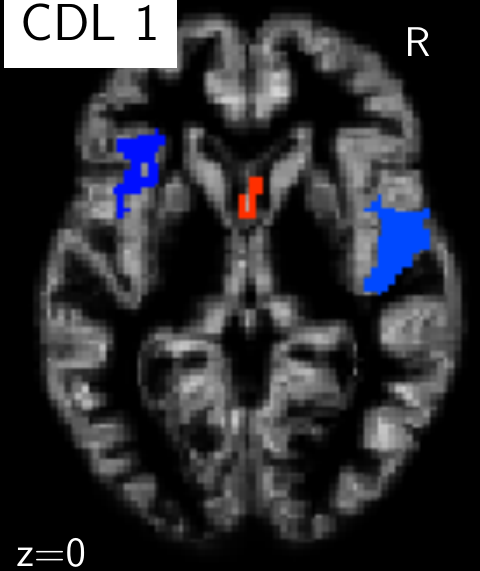}
    \end{minipage}
    \begin{minipage}[c]{.18\linewidth}
        \centering
        \includegraphics[width=\linewidth]
        {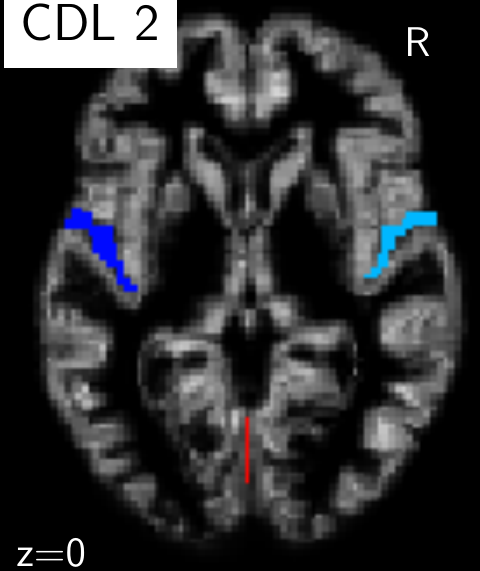}
    \end{minipage}
    \begin{minipage}[c]{.18\linewidth}
        \centering
        \includegraphics[width=\linewidth]
        {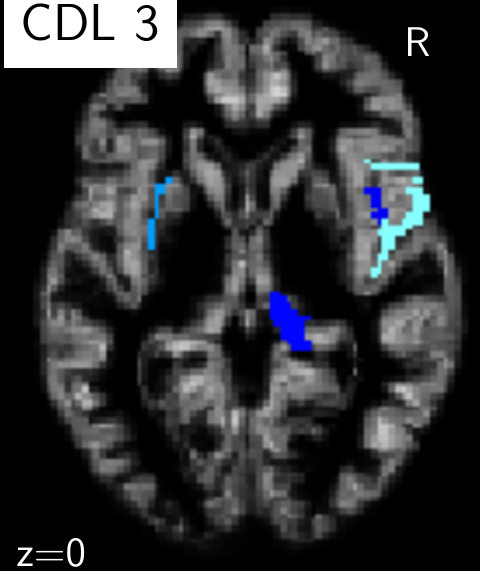}
    \end{minipage}
    \hspace{5mm}
    \begin{minipage}[c]{.27\linewidth}
        \centering
        \includegraphics[width=\linewidth]
        {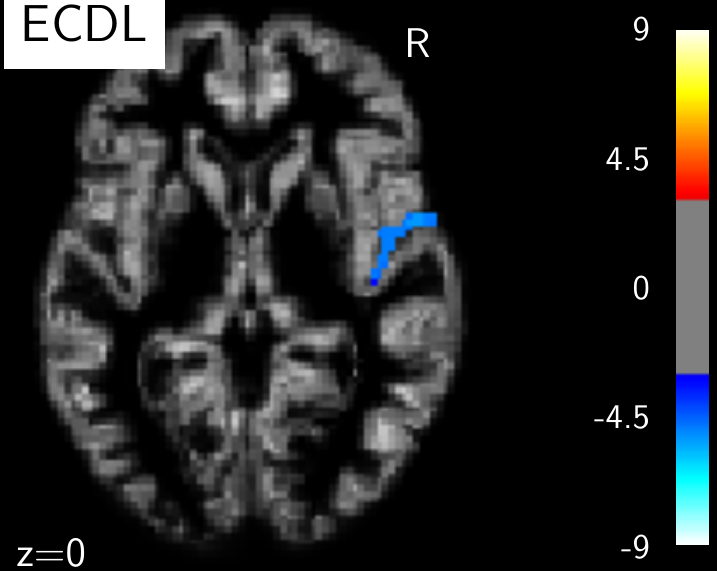}
    \end{minipage}
    \caption{\label{fig:oasis_haxby}
      Results of the CDL and ECDL algorithms on Haxby (top) and Oasis (bottom) experiments. CDL algorithm outcomes are highly dependent on the clustering, which creates a jitter in the solution. Drawing consensus among many CDL results, ECDL removes the arbitrariness related to the clustering scheme.
      }
\end{figure}
\paragraph{\textbf{Results.}}
The results of these experiments are displayed in \Cref{fig:oasis_haxby} with Z-transform of the p-values.
For clarity, we thresholded the Z-score maps at $3$ (and $-3$) keeping only the regions that have a high probability of being discriminative.
The solutions given by the CDL algorithm with three different choices of clustering look noisy and unstable while the ECDL solution defines a synthesis of the CDL results and exhibits a nice symmetry in the case of Haxby.
Thus, these results clearly illustrate that the ensembling step removes the arbitrariness due to the clustering.
\paragraph{\textbf{Stability of Bagging Estimator.}}
\begin{figure}[tb]
    \centering
    \begin{minipage}[c]{.41\linewidth}
        \centering
	    \includegraphics[width=\linewidth]
    	{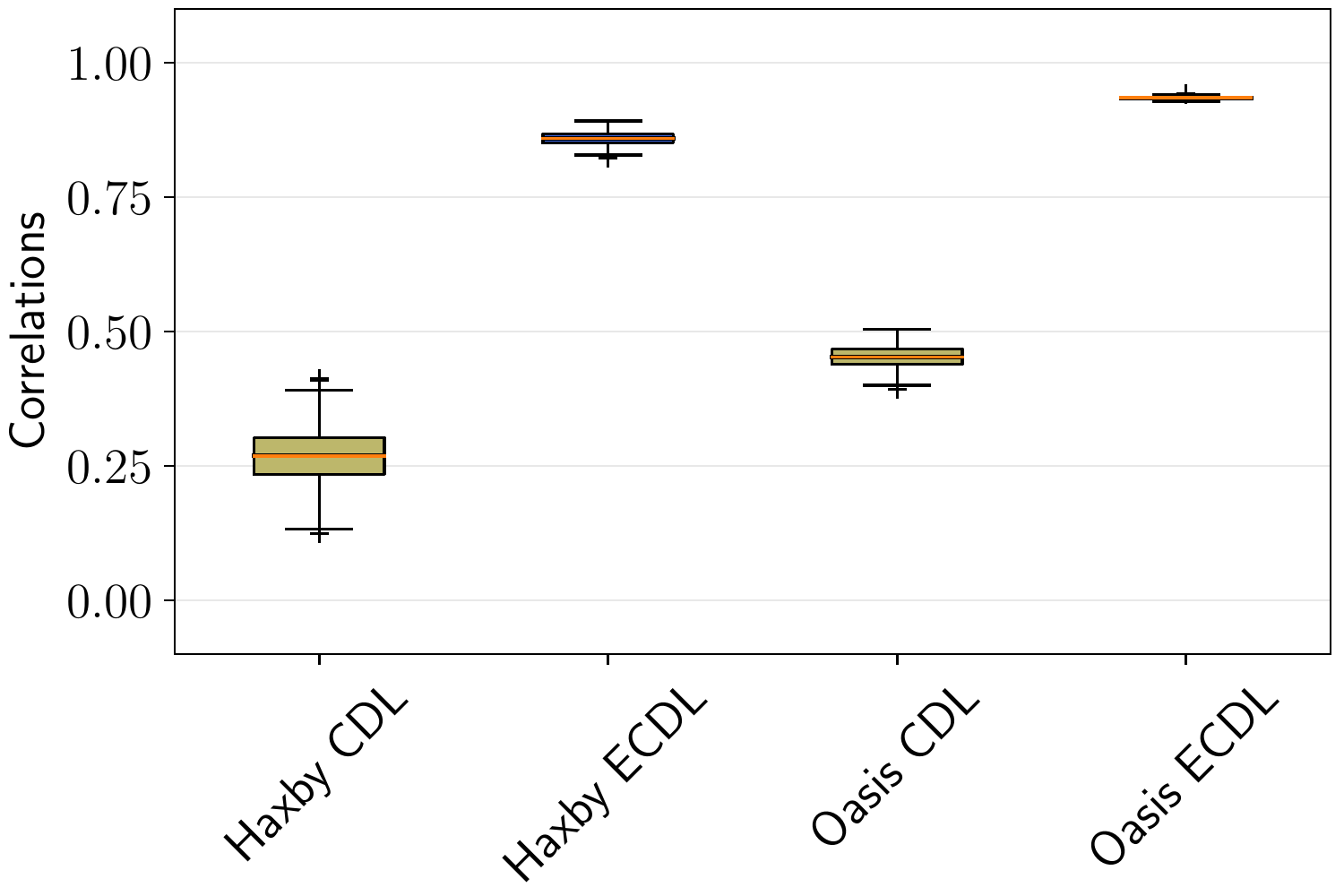}
    \end{minipage}
    \begin{minipage}[c]{.41\linewidth}
        \centering
        \includegraphics[width=\linewidth]
        {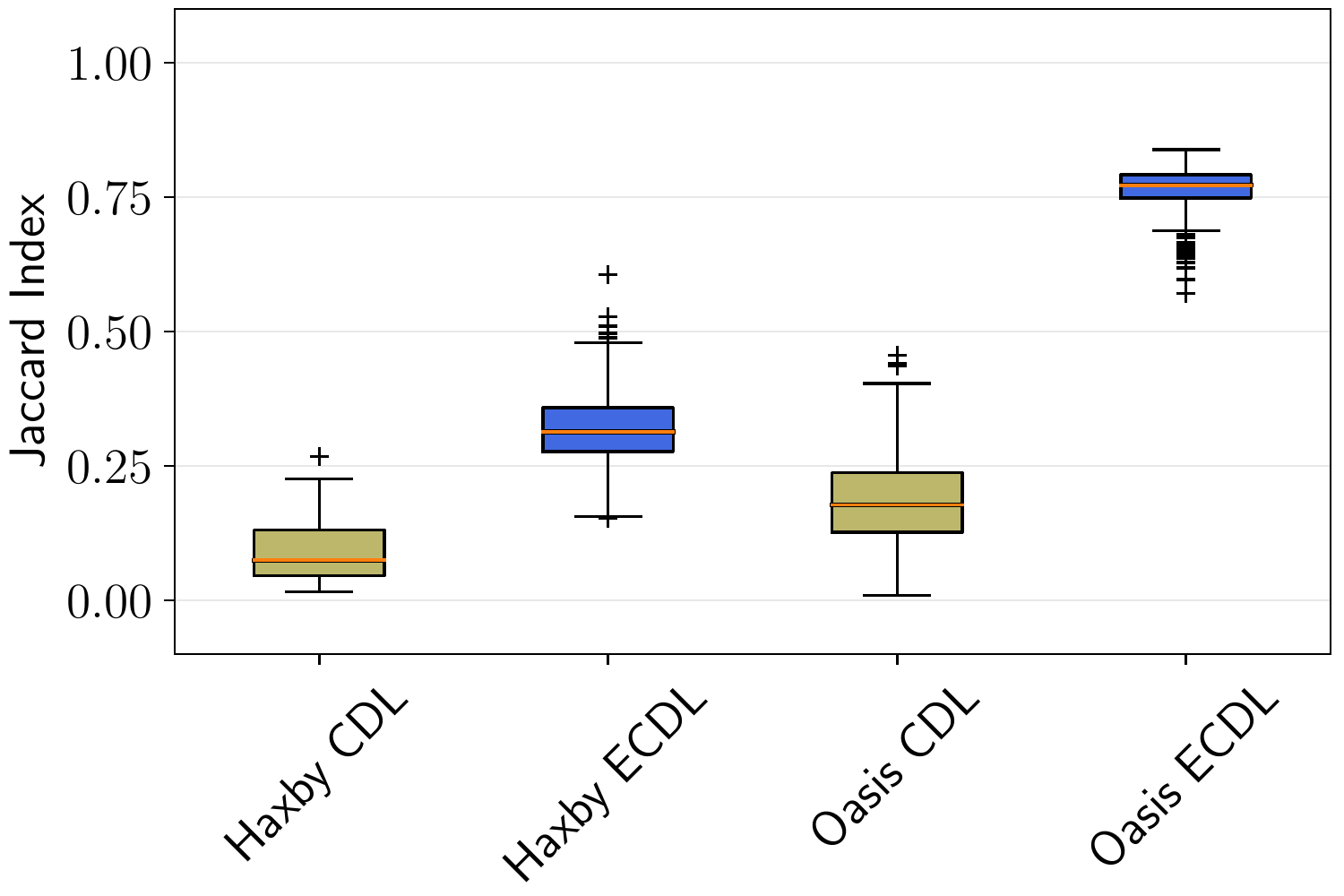}
    \end{minipage}
    \caption{\label{fig:stability} Correlation (left) and Jaccard
      Index (right) are much higher with the ECDL algorithm than with
      CDL across 25 replications of the analysis of the imaging datasets.}
\end{figure}
This last experiment quantifies the gain in stability when adding the ensembling step.
From the two previous experiments, we derive $25$ ECDL solution maps (with $B = 25$) and $25$ CDL solution maps and measure the variability of the results.
Correlation between the full maps and Jaccard index of the detected areas (here, voxels with an absolute Z-score greater than $3$) show that ECDL is substantially more stable than CDL.
%
%%%%%%%%%%%%%%%%%%%%%%%%%%%%%%%%%%%%%%%%%%%%%%%%%%%%%%%%%%%%%%%%%%%%%%%%%%%%%%%
%%%%%%%%%%%%%%%%%%%%%%%%%%%%%%%%%%%%%%%%%%%%%%%%%%%%%%%%%%%%%%%%%%%%%%%%%%%%%%%
\section{Discussion}
\label{sec:discussion}
%%%%%%%%%%%%%%%%%%%%%%%%%%%%%%%%%%%%%%%%%%%%%%%%%%%%%%%%%%%%%%%%%%%%%%%%%%%%%%%
%%%%%%%%%%%%%%%%%%%%%%%%%%%%%%%%%%%%%%%%%%%%%%%%%%%%%%%%%%%%%%%%%%%%%%%%%%%%%%%
%
\paragraph{\textbf{Recapitulation.}} We have introduced ECDL, an algorithm for high-dimensional inference on structured data which scales even when the number of predictors $p \geq 10^{5}$ is much higher than the number of samples
$n \leq 10^{3}$.
It can be summarized as follows: i) perform $B$ repetitions of the CDL algorithm, that runs \DL (DL) inference on a model compressed by clustering,  yielding several p-values for each predictor; ii) use an ensemble method aggregating all p-values to derive a single p-value per predictor.
In \Cref{sub:first_simu_clustering}, we have shown that the clustering step,  justified by specific data structures and locally high inter-predictor correlation, was necessary to yield an informative inference solution when $n \ll p$.
Then, we have demonstrated, in \Cref{sub:second_simu_boostrap_aggregation}, that randomizing and bagging the CDL solution improves the control of the FWER and the precision-recall curve.
While the ensembling step obviously removes the arbitrariness of the clustering choice, in \Cref{sub:experiments_mri} we showed it also increases stability.
\paragraph{\textbf{ECDL Parameter Setting.}}
The number of clusters $C$ is the main free parameter, and an optimal value depends on characteristics  of the data (inter-predictor correlation, SNR).
In our simulations, we observe a bias/variance trade-off: a small number of clusters reduces variance and enhances statistical power, while a greater number yields refined solutions.
The ensembling step helps improving shape accuracy without loss in sensitivity, as the combination of multiple CDL solutions recovers finer spatial information.
%
%We propose to rely on $C=500$ for brain imaging.
% Indeed the number of clusters considered in ECDL is equal to $B \times C$ giving more granularity to the solution.
%
\paragraph{\textbf{Computational Cost of ECDL.}}
The most expensive step is the DL inference, which includes the resolution of $\cO(C)$ Lasso problems with $n$ samples and $C$ features.
This is repeated $B$ times in ECDL, making it embarrassingly parallel; we could run the ECDL algorithm on standard desktop stations with $n = 400$, $C = 500$ and $B = 25$ in less than 10 minutes.
\paragraph{\textbf{Additional Work Related to ECDL.}}
In Haxby experiment (\lcf \Cref{sub:experiments_mri}) we approximated the problem as a regression one.
Thus, an interesting extension would be to adapt ECDL to classification settings.
Another matter is the comparison with bootstrap and permutation-based approaches \eg \cite{Gaonkar2012}, that we leave for future work.
Note however that, in \cite{Dezeure2015}, a study of bootstrap approaches points out some unwanted properties and they do not outperform DL.
\paragraph{\textbf{Usefulness for Medical Imaging.}}
For structured high-dimensional data, our algorithm is relevant to assess the statistical significance of a set of predictors when fitting a target variable.
Our experimental results show that ECDL is very promising for inference problems in medical imaging.
\paragraph{\textbf{Acknowledgements.}}
This research was supported by Labex DigiCosme (project ANR-11-LABEX-0045-DIGICOSME) operated by ANR as part of the program
"Investissement d'Avenir" Idex Paris Saclay (ANR-11-IDEX-0003-02).
This work is also funded by the FAST-BIG project (ANR-17-CE23-0011).
\bibliographystyle{plain}
\bibliography{biblio}
\end{document}